\documentstyle[12pt,epsf]{article}
\begin{document}
\begin{titlepage}
\begin{flushright}
Alberta-Thy-19-95 \\
gr-qc/9510040
\end{flushright}
\vfill

\begin{center}
{\Large \bf   Covariant double-null dynamics: $(2+2)$-splitting
of the Einstein equations.}\\
\vfill P. R. Brady$^{\dag}$, S. Droz$^{\ddag}$, W. Israel$^{\ddag}$ and
S. M. Morsink$^{\ddag}$.
\vspace{2cm}\\
$^{\dag}$Theoretical Astrophysics 130-33, California
Institute of Technology, Pasadena, CA 91125, U.S.A.

$^{\ddag}$Canadian Institute for Advanced Research Cosmology Program,
Theoretical Physics Institute, University of Alberta, Edmonton,
Canada T6G 2J1
\vfill
%%%%%%%%%%%%%%%%%%%%%%%%%%%%%%%%%%%%%%%%%%%%%%%%
%
%ABSTRACT
%
%%%%%%%%%%%%%%%%%%%%%%%%%%%%%%%%%%%%%%%%%%%%%%%%

\begin{abstract}
 The paper develops a $(2+2)$-imbedding formalism adapted to
a double foliation of spacetime by a net of two
intersecting families of lightlike hypersurfaces. The
formalism is two-dimensionally covariant, and
leads to simple, geometrically transparent and
tractable expressions for the Einstein field equations and
the Einstein-Hilbert action, and it should find a variety of
applications. It is applied here to elucidate the structure
of the characteristic initial-value problem of general
relativity.\end{abstract}
%{To appear: ?????}}
\vfill\vfill
\end{center}
\end{titlepage}

\newpage
\section{Introduction}
The classic analysis of Arnowitt, Deser and Misner (ADM) \cite{ref:1}
formulates gravitational dynamics in terms of the evolution of a
spatial 3-geometry. The geometrical framework is the imbedding
formalism of Gauss and Codazzi for the foliation of spacetime by
spacelike hypersurfaces \cite{ref:2}.

Quite often, however, one encounters circumstances where a
lightlike
foliation is especially suitable. Because of the degeneracies that
arise in
the lightlike case the imbedding relations are very different and
the
situation not quite so familiar and under control. To bypass
the degeneracies, one is forced to fall back to a foliation of
codimension~2, by spacelike 2-surfaces. It is our aim in this paper
to develop a simple $(2+2)$-imbedding formalism of this kind.

Several $(2+2)$-formalisms are extant \cite{ref:3}, the earliest and best
known being the generalized spin-coefficient formalism of Geroch,
Held and Penrose (GHP) \cite{ref:4}. Basically, of course, all such
formalisms
have the same content, but they take very different forms.

The essential feature of the present approach is that it maintains
manifest two-dimensional  covariance while
operating with objects having direct geometrical meaning.
Two-dimensional covariance permits reduction of the Einstein field
equations to an especially concise and transparent form: the ten
Ricci components are embraced in a set of just three compact,
two-dimensionally covariant expressions.

There is a limitation, at least in the version presented here. (It
applies to most of the formalisms we have listed \cite{ref:3}.) The two
independent normals to an imbedded 2-surface---conveniently taken
as
a pair of lightlike vectors, since their directions are uniquely
defined---are assumed from the beginning to be
hypersurface-orthogonal. This precludes choosing them as principal
null vectors of the Weyl tensor for a twisting geometry like Kerr.
In
this respect, the formalism is less flexible than GHP, and not as
well tailored for the study of algebraically special metrics.

$(2+2)$ formalisms have a wide range of applications: to the analysis
of the characteristic initial-value problem \cite{ref:5}, the dynamics
of strings \cite{ref:6} and of real and apparent horizons \cite{ref:7}
and light-cone quantization \cite{ref:8} and gravitational interactions in
ultra-high energy collisions \cite{ref:new9}.   In a separate publication
\cite{ref:9}, we shall use the present formalism to study the nature
of the singularity at the Cauchy horizon in a generic black hole.

We conclude this Introduction by briefly outlining the contents of the
paper.  The basic metrical notions (adapted co-ordinates, basis
vectors and form of the metric) are defined in Sec.~2.  In Sec.~3, we
introduce in two-dimensionally covariant form the geometrical
information encoded in first derivatives of the metric: the extrinsic
curvatures and ``twist,'' as well as the invariant operators which
perform differentiation along the two lightlike normals.  This
comprises the basic formal machinery needed in Sec.~4, which presents
the central result of the paper, the tetrad components of the Ricci
tensor as three concise equations (\ref{eq:27})--(\ref{eq:29}).  (To
make direct access to these results easy, their derivation is deferred
to the second half of the paper (Secs.~9--12), which also provides
(Sec.~13) the tetrad components of the full Riemann tensor.)

The contracted Bianchi identities (Sec.~5) are applied in Sec.~7 to
analyze the structure of the characteristic initial-value problem.  In
Sec.~8 we sketch the Lagrangian formulation of covariant double-null
dynamics.

The Ricci and Riemann components result from the commutation relations
for four-di\-men\-sio\-nal covariant differentiation.  Their most
efficient derivation calls for a formalism that is both four- and
two-dimensionally covariant.  Unfortunately, these two requirements do
not mesh easily.  Four-dimensional covariance tends to clutter the
formulae by treating subsidiary two-dimensional quantities like shift
vectors and the two-dimensional connection as 4-scalars, on a par with
the primary geometrical properties, extrinsic curvature and twist.
Those properties, for their part, are correlated, not with
four-dimensional covariant derivatives, but with Lie derivatives,
which are non-metric and have no direct link to curvature.  To patch
up these differences, and thus streamline the derivations, seems to
need a certain degree of artifice.  In Sec.~10 we address this (purely
technical) problem by temporarily working with a ``rationalized''
covariant derivative which exhibits both four-dimensional and
restricted (``rigid'') two-dimensional covariance.

Some brief remarks (Sec.~14) conclude the paper.

\section{$(2+2)$-split of the metric}
We shall suppose that we are given a foliation of spacetime by
lightlike hypersurfaces $\Sigma^0$ with normal generators
$\ell_\alpha^{(0)}$,
and a second, independent foliation by lightlike hypersurfaces
$\Sigma^1$ with generators $\ell_\alpha^{(1)}$ nowhere parallel to
$\ell_\alpha^{(0)}$. The intersections of $\{\Sigma^0\}$ and $\{\Sigma^1\}$
define a foliation of codimension 2 by spacelike 2-surfaces $S$.
(The
topology of $S$ is unspecified. All our considerations are local.)
$S$ has exactly two lightlike normals at each of its points,
co-directed with $\ell^{(0)}$ and $\ell^{(1)}$.

In terms of local charts, the foliation is described by the
imbedding
relations
\begin{equation}
x^\alpha=x^\alpha(u^A,\theta^a).
\label{eq:1}
\end{equation}
Here, $x^\alpha$ are four-dimensional spacetime co-ordinates (assumed
admissible in the sense of Lichnerowicz \cite{ref:10}); $u^0$ and $u^1$ are
a
pair of scalar fields constant over each of the hypersurfaces
$\Sigma^0$ and $\Sigma^1$ respectively; and $\theta^2$, $\theta^3$ are
intrinsic co-ordinates of the 2-spaces $S$, each characterized by
a
fixed pair of values $(u^0,u^1)$.

{\bf Notation:} Our conventions are: Greek indices $\alpha,\beta,\dots$
run from 0 to 3; upper-case Latin indices $A,B,\dots$ take values
$(0,1)$; and lower-case Latin indices $a,b,\dots$ take values
$(2,3)$. We adopt MTW curvature conventions \cite{ref:2} with signature
$(-+++)$ for the spacetime metric $g_{\alpha\beta}$. When there is no
risk
of confusion we shall often omit the Greek indices on 4-vectors
like
$\ell_\alpha^{(A)}$ and $e_{(a)}^\alpha$: they are easily identifiable as
4-vectors by their parenthesized labels. Four-dimensional covariant
differentiation is indicated either by $\nabla_\alpha$ or a vertical
stroke: $\nabla_\beta A_\alpha\equiv A_{\alpha\mid\beta}$. Four-dimensional
scalar
products are often indicated by a dot: thus,
$\ell_{(A)}\cdot\ell_{(B)}\equiv
g_{\alpha\beta}\,\ell_{(A)}^\alpha\,\ell_{(B)}^\beta$.
Further conventions will be introduced as the need arises.

Without essential loss of generality we may assume the functions
$x^\alpha(u^A,\theta^a)$ to be smooth (at least thrice differentiable).
(We
are always free to make the co-ordinate choice $x^A=u^A$,
$x^a=\theta^a$, but at the cost of losing manifest four-dimensional
and
two-dimensional covariance.)

The lightlike character of the hypersurfaces $\Sigma^A$ is encoded in
\begin{equation}
\nabla u^A\cdot \nabla u^B\equiv g^{\alpha\beta}
(\partial_\alpha u^A)(\partial_\beta u^B)=e^{-\lambda}
\eta^{AB}
\label{eq:2}
\end{equation}
for some scalar field $\lambda(x^\alpha)$, where
\begin{equation}
\eta^{AB}=\mbox{\rm anti-diag}\,(-1,-1)=\eta_{AB};
\label{eq:3}
\end{equation}
$\eta^{AB}$ and its inverse $\eta_{AB}$ are employed to raise and
lower upper-case Latin indices,
e.g., $\ell_{(0)}=-\ell^{(1)}$.

The generators $\ell^{(A)}$ of $\Sigma^A$ are parallel to the
gradients of $u^A(x^\alpha)$.  It is symmetrical and convenient to
define $\ell_{(A)}=e^\lambda\nabla u^A$, i.e.,
\begin{equation}
\ell_\alpha^{(A)}= e^\lambda\partial_\alpha u^A.
\label{eq:4}
\end{equation}
Then
\begin{equation}
\ell_{(A)}\cdot\ell^{(B)}=e^\lambda \delta_A^B.
\label{eq:5}
\end{equation}

The pair of vectors $e_{(a)}$, defined from (\ref{eq:1}) by
\begin{equation}
e_{(a)}^\alpha=\partial x^\alpha/\partial\theta^a,
\label{eq:6}
\end{equation}
are holonomic basis vectors tangent to $S$. The intrinsic metric
$g_{ab}\,d\theta^a\, d\theta^b$ of $S$ is determined by their scalar
products:
\begin{equation}
g_{ab}=e_{(a)}\cdot e_{(b)}.
\label{eq:7}
\end{equation}
Lower-case Latin indices are lowered and raised with $g_{ab}$ and
its
inverse $g^{ab}$; thus $e^{(a)}\equiv g^{ab}e_{(b)}$ are the dual
basis vectors tangent to $S$, with $e^{(a)}\cdot e
_{(b)}=\delta_b^a$.
Since $\ell^{(A)}$ is normal to every vector in $\Sigma ^A$, we have
\begin{equation}
\ell^{(A)}\cdot e_{(a)}=0.
\label{eq:8}
\end{equation}

\begin{center}
  \mbox{
     \epsfxsize=16cm
     \epsffile{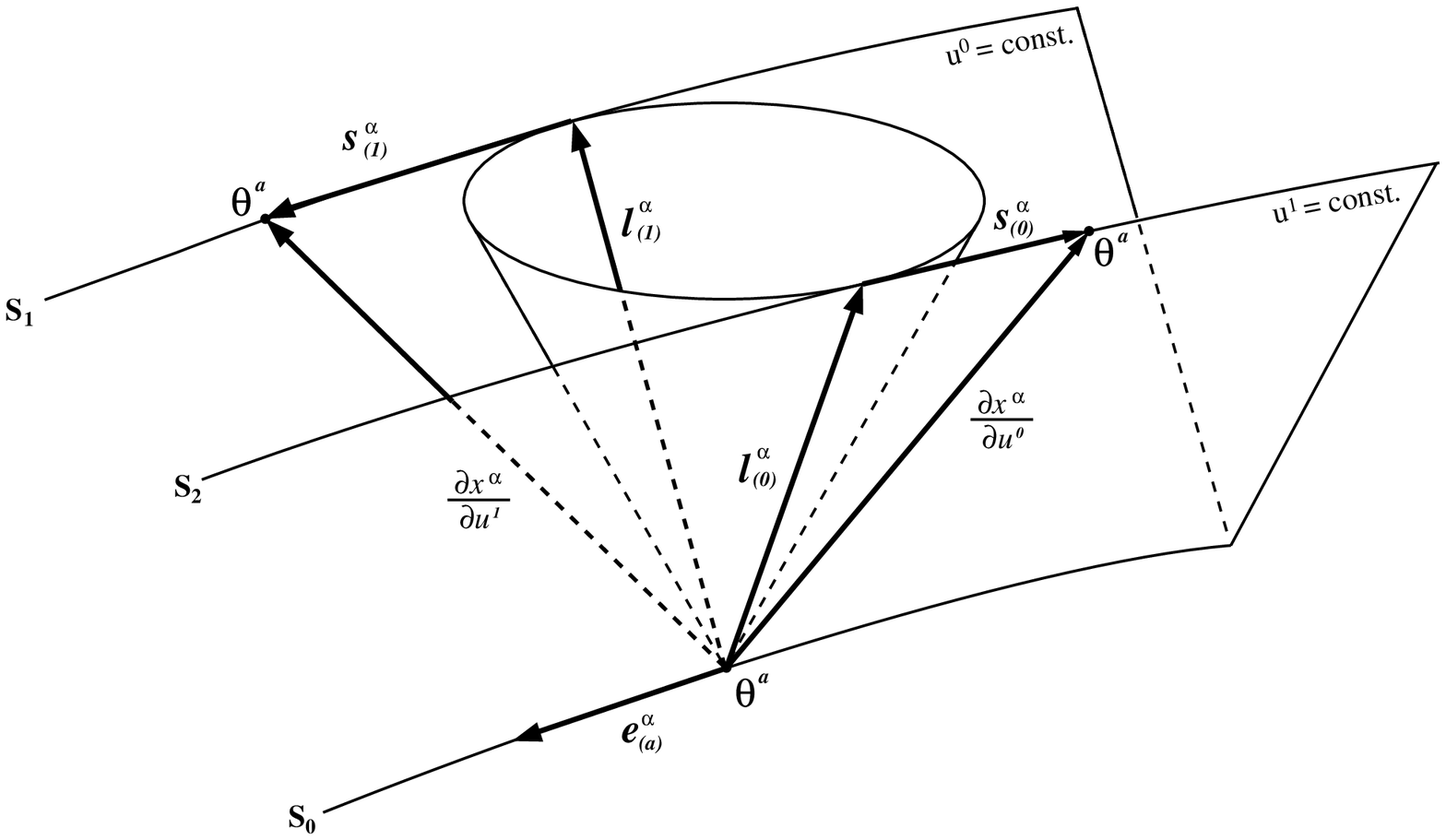}
     }
     \begin{quote}
      Figure 1: The $2+2$ splitting of the four dimensional
      space time into a foliation of intersecting null surfaces $u^0=const.$
      and $u^1=const.$.
      \end{quote}
\end{center}

In general, $\theta^a$ cannot be chosen so as to remain constant along
{\it both\/} sets of generators $\ell^{(A)}$. They are convected
(Lie-transported) along the pair of vector fields $\partial x^\alpha/\partial
u^A$
(in general, non-lightlike).

{}From (\ref{eq:4}) and (\ref{eq:5}) one finds that $\partial
x^\alpha/\partial u^A-\ell_{(A)}^\alpha$ is orthogonal to
$\ell_{(B)}$, i.e., tangent to $S$.  This validates the decomposition
\begin{equation}
  \frac{\partial x^\alpha}{\partial u^A}=\ell_{(A)}^\alpha+s_A^a
  \,e_{(a)}^\alpha,
  \label{eq:9}
\end{equation}
thus defining a pair of ``shift vectors'' $s_A^\alpha$ tangent to $S$
(see Fig.~1).

An arbitrary displacement $dx^\alpha$ in spacetime is, according to
(\ref{eq:6}) and (\ref{eq:9}), decomposable as
\begin{equation}
dx^\alpha=\ell_{(A)}^\alpha\,du^A+e_{(a)}^\alpha(d\theta^a+s_A^a\,du^A).
\label{eq:10}
\end{equation}

{}From (\ref{eq:5}), (\ref{eq:7}) and (\ref{eq:8}) we read off the
completeness relation
\begin{equation}
g_{\alpha\beta}=e^{-\lambda}\eta_{AB}\ell_\alpha^{(A)}\ell_\beta^{(B)}+g_{ab}
e_\alpha^{(a)}\,e_\beta^{(b)}.
\label{eq:11}
\end{equation}
Combining (\ref{eq:10}) and (\ref{eq:11}) shows that the spacetime metric is
decomposable as
\begin{equation}
	g_{\alpha\beta}dx^\alpha\,dx^\beta=e^\lambda\eta_{AB}
	\,du^A\,du^B+g_{ab}(d\theta^a+s_A^a\,du^A)
	(d\theta^b+s_B^b\,du^B).
	\label{eq:12}
\end{equation}

\section{Two-dimensionally covariant objects embodying
first derivatives of the metric: extrinsic curvatures $K_{Aab}$,
twist $\omega^a$ and normal Lie derivatives $D_A$}
Absolute derivatives of four-dimensional tensor fields with respect
to $u^A$ and $\theta^a$ are projections of the four-dimensional
covariant derivative $\nabla_\alpha$, and denoted by
\begin{equation}
\frac{\delta}{\delta u^A}=
\frac{\partial x^\alpha}{\partial u^A}
\,\nabla_\alpha,\qquad\frac{\delta}{\delta\theta^a}=e_{(a)}
\cdot\nabla.
\label{eq:13}
\end{equation}

{}From (\ref{eq:6}) and the symmetry of the mixed partial derivatives and the
affine connection,
\begin{equation}
\frac{\delta e_{(a)}^\alpha}{\delta u^A}=\frac{\delta}{\delta\theta^a}
\left(\frac{\partial x^\alpha}{\partial u^A}\right),\qquad
\frac{\delta e_{(a)}}{\delta\theta^b}=\frac{\delta e_{(b)}}{\delta\theta^a}.
\label{eq:14}
\end{equation}

The object
\begin{equation}
\Gamma_{ab}^c=e^{(c)}\cdot \delta e_{(a)}/\delta\theta^b
\label{eq:15}
\end{equation}
is, as the notation suggests, the Christoffel symbol associated
with
$g_{ab}$, as is easily verified by forming $\partial_c g_{ab}$,
recalling
(\ref{eq:7}) and applying Leibnitz's rule.

Associated with its two normals $\ell_{(A)}$, $S$ has two extrinsic
curvatures $K_{Aab}$, defined by
\begin{equation}
	K_{Aab}=e_{(a)}\cdot \delta \ell_{(A)}
	/\delta\theta^b=\ell_{(A)\alpha\mid\beta}
	\,e_{(a)}^\alpha\,e_{(b)}^\beta.
	\label{eq:16}
\end{equation}
(Since we are free to rescale the null vectors $\ell_{(A)}$, a
certain scale-arbitrariness is inherent in this definition.)
Because
of (\ref{eq:8}), we can rewrite
\begin{equation}
K_{Aab}=-\ell_{(A)}\cdot \delta e_{(a)}/\delta\theta^b,
\label{eq:17}
\end{equation}
which exhibits the symmetry in $a$, $b$.

A further basic geometrical property of the double
foliation is given by
the Lie bracket of $\ell
_{(B)}$ and $\ell_{(A)}$, i.e., the 4-vector
\begin{equation}
[\ell_{(B)},\ell_{(A)}]^\alpha=2(\ell_{[(B)}\cdot\nabla)\ell_{(A)]}^\alpha.
\label{eq:18}
\end{equation}
Noting (\ref{eq:9}) and the fact that the Lie bracket of the vectors
$\partial x^\alpha/\partial u^B$ and $\partial x^\alpha/\partial u^A$
vanishes identically, and recalling (\ref{eq:14}), we find
\begin{equation}
  [\ell_{(B)},\ell_{(A)}]=\epsilon_{AB}\,\omega^a\,e_{(a)},
  \label{eq:19}
\end{equation}
where the 2-vector $\omega^a$ is given by
\begin{equation}
\omega^a=\epsilon^{AB} (\partial_B s_A^a-s_B^b s_{A;b}^a),
\label{eq:20}
\end{equation}
the
semicolon indicates
two-dimensional covariant  differentiation
associated with metric $g_{ab}$, and $\epsilon_{AB}$ is the
two-dimensional permutation symbol, with $\epsilon_{01}=+1$.
(Note that raising indices with $\eta^{AB}$ to form
$\epsilon^{AB}$ yields $\epsilon^{10}=+1$.)

The geometrical significance of the ``twist'' $\omega^a$ can
be read off from (\ref{eq:19}): the
curves tangent to the
generators  $\ell_{(0)},\,\ell_{(1)}$ mesh together to form
2-surfaces (orthogonal to the surfaces $S$) if and only if
$\omega^a=0$.
In this case, it would be consistent to allow the co-ordinates
$\theta^a$ to be dragged along both sets of generators, and thus to
gauge both shift vectors to zero.

We denote by $D_A$ the two-dimensionally invariant operator
associated with differentiation along the normal direction
$\ell_{(A)}$. Acting on any two-dimensional geometrical
object $X_{\ b\dots}^{a\dots}$, $D_A$ is formally defined
by
\begin{equation}
D_A X^{a\dots}_{\ b\dots}=(\partial_A-{\cal L}_{s_A^d})X_{\
b\dots}^{a\dots}.
\label{eq:21}
\end{equation}
Here, $\partial_A$ is the partial derivative with respect to
$u^A$ and ${\cal L}_{s_A^d}$ the Lie derivative with respect to
the 2-vector $s_A^d$.

As examples of (\ref{eq:21}), we have for a 2-scalar $f$ (this
includes any object bearing upper-case, but no lower-case,
Latin indices):
\begin{equation}
D_Af=(\partial_A-s_A^a \partial_a)f=\ell_{(A)}^\alpha\partial_\alpha f
\label{eq:22}
\end{equation}
(in which the second equality follows at once from (\ref{eq:9})); and
for the 2-metric $g_{ab}$:
\begin{equation}
D_A g_{ab}=\partial_A g_{ab}-2s_{A(a;b)}=2K_{Aab},
\label{eq:23}
\end{equation}
in which the second equality is derivable from (\ref{eq:7}), (\ref{eq:14})
and (\ref{eq:9}). (For the detailed derivation, see (\ref{eq:75}) below,
or Appendix~B.)

The geometrical meaning of $D_A$ is quite generally the
following (see Appendix~B):\newline
$D_A X_{\ b\dots}^{a\dots}$ is the
projection onto $S$ of the Lie derivative of the equivalent
tangential 4-tensor
\[
X_{\ \beta\dots}^{\alpha\dots}\equiv X_{\ b\dots}^{a\dots}
e_{(a)}^\alpha e_\beta^{(b)}\dots
\]
with respect to the 4-vector $\ell_{(A)}^\mu$:
\begin{equation}
  D_A X_{\ b\dots}^{a\dots}=e_\alpha^{(a)} e_{(b)}^\beta\cdot
 {\cal L}_{\ell_{(A)}^\mu} X_{\beta\dots}^{\alpha\dots}\;.
\label{eq:24}
\end{equation}

The objects $K_{Aab}$, $\omega^a$ and $D_A$ comprise all the
geometrical structure that is needed for a succinct
two-dimensionally covariant expression of the Riemann and
Ricci curvatures of spacetime. According to (\ref{eq:16}),
(\ref{eq:19}) and
(\ref{eq:24}), all are simple projections onto $S$ of
four-dimensional geometrical objects. Consequently, they
transform very simply under two-dimensional co-ordinate
transformations. Under the arbitrary reparametrization
\begin{equation}
\theta^a\to \theta^{a'}=f^a(\theta^b, u^A)
\label{eq:25}
\end{equation}
(which leaves $u^A$ and hence the surfaces $\Sigma^A$ and $S$
unchanged), $\omega_a$ and $K_{Aab}$ transform cogrediently
with
\begin{equation}
e_{(a)}\to e_{(a)}'=e_{(b)} \partial\theta^b/\partial\theta^{a'}
\label{eq:26}
\end{equation}
(see (\ref{eq:6})), and $D_A$ is invariant.  By contrast, $\partial
x^\alpha/\partial u^A$ and hence the shift vectors $s_A^a$ (see
(\ref{eq:9})) undergo a more complicated gauge-like transformation,
arising from the $u$-dependence in (\ref{eq:25}).

\section{Ricci tensor}
The geometrical and notational groundwork laid in the
previous sections allows us now to simply display the
components of the Ricci tensor, deferring derivations to
Secs.~9--12. Our notation for the tetrad components is
typified by
\[
^{(4)}\! R_{ab}=R_{\alpha\beta}e_{(a)}^\alpha e_{(b)}^\beta,\qquad
R_{aA}=R_{\alpha\beta} e_{(a)}^\alpha \ell_{(A)}^\beta.
\]

The results are
\begin{eqnarray}
%R_ab
  ^{(4)}\!R_{ab}
  &=&\frac12\,^{(2)}\!Rg_{ab}-e^{-\lambda}
  \left(D_A+K_A\right)K^A_{ab} \nonumber \\
  & &\qquad +2e^{-\lambda}K_{A(a}^{\;\;\;d} K_{b)d}^A-\frac12
  e^{-2\lambda}\,{\omega_a\omega_b}-\lambda_{;ab}-
  \frac12\lambda_{,a}\lambda_{,b}
  \label{eq:27}\\
%R_AB
  R_{AB}&=&
  -D_{(A}K_{B)}-K_{Aab} K_B\,^{ab}+K_{(A}D_{B)}\lambda \nonumber\\
  & &\qquad -\frac12\eta_{AB}
  \left[
    \left(D^E+K^E\right)D_E\lambda-e^{-\lambda}
    \omega^a\omega_a+(e^\lambda)^{;a}\,_a
  \right]
  \label{eq:28}\\
%R_Aa
  R_{Aa}&=&K_{Aa;b}^{\;b}-\partial_a
  K_A-\frac12\partial_aD_A\lambda+\frac12K_A\partial_a\lambda \nonumber\\
  & &\qquad +\frac12 \epsilon_{AB}e^{-\lambda}
  \left[
    \left(D^B+K^B\right)\omega_a-\omega_a D^B \lambda
  \right],
  \label{eq:29}
\end{eqnarray}
where $^{(2)}\!R$ is the curvature scalar associated with
the 2-metric $g_{ab}$, and $K_A\equiv K_{A\ a}^{\ a}$.

\section{Bianchi identities. Bondi's lemma}
The Ricci components are linked by four differential
identities, the contracted Bianchi identities
\begin{equation}
\nabla_\beta R_{\ \alpha}^\beta=\frac12 \partial_\alpha R,
\label{eq:30}
\end{equation}
where the four-dimensional curvature scalar
$R=R_{\ \alpha}^\alpha$ is given by
\begin{equation}
R=e^{-\lambda} R_{\ A}^A+R_{\ a}^a,
\label{eq:31}
\end{equation}
according to (\ref{eq:11}).

As we show in Sec.~12,
projecting (\ref{eq:30}) onto $e_{(a)}$ leads to
\begin{equation}
\left(D_A+K_A\right) R^A{}_a=\frac12 \partial_a R^A{}_A+\frac12 e^\lambda
\partial_a\,^{(4)}\!R^b{}_b-\left(e^\lambda\,^{(4)}\!R^b{}_a\right)_{;b}.
\label{eq:32}
\end{equation}

Projection of (\ref{eq:30}) onto $\ell_{(A)}$ similarly yields
\begin{eqnarray}
\lefteqn{
D_B\left(R_{\ A}^B-\frac12\delta_A^B R_{\ D}^D\right)-\frac12 e^\lambda D_A
\,^{(4)}\!R_{\ a}^a} \nonumber \\
& &=e^\lambda \,^{(4)}\!R_{ab} K_A^{\ ab}-R_{\ A}^B K_B-\left(e^\lambda
R_{\ A}^a\right)_{;a}+\epsilon_{AB} \omega^a R_{\ a}^B.
\label{eq:33}
\end{eqnarray}

Equations (\ref{eq:32}) and (\ref{eq:33}) express the four Bianchi
identities in terms of the tetrad components of the Ricci tensor.

We now look at the general structure of these equations.

For $A=0$ in (\ref{eq:33}), $R_0^0$ does not contribute to the first
(parenthesized) term, since
\begin{equation}
-R_{01}=R_0^0=R_1^1=\frac12 R_A^A.
\label{eq:34}
\end{equation}
This equation therefore takes the form
\begin{equation}
D_1 R_{00}+\frac12 e^\lambda D_0\,^{(4)}\!R_a^a=-K_0 R_{01}+
{\cal L}(^{(4)}\!R_{ab},R_{00},R_{0a},\partial_a),
\label{eq:35}
\end{equation}
in which the schematic notation ${\cal L}$ implies that the
expression is linear homogeneous in the indicated Ricci
components and their two-dimensional spatial derivatives
$\partial_a$.

The other $(A=1)$ component of (\ref{eq:33}) has the analogous
structure
\begin{equation}
D_0 R_{11}+\frac12 e^\lambda D_1 \,^{(4)}\!R_a^a=-K_1 R_{01}+
{\cal L}(^{(4)}\!R_{ab},R_{11}, R_{1a},\partial_a).
\label{eq:36}
\end{equation}

The form of the remaining two Bianchi identities (\ref{eq:32}) is
\begin{equation}
D_0 R_{1a}+D_1 R_{0a}={\cal L}(^{(4)}\!R_{ab},R_{01}, R_{Aa},
\partial_a).
\label{eq:37}
\end{equation}

It is noteworthy that the appearance of $R_{01}$ in (\ref{eq:35}) and
(\ref{eq:36}) is purely algebraic: its vanishing would be a direct
consequence of the vanishing of just six of the other components.
Bondi et al \cite{ref:12} and Sachs \cite{ref:5} therefore refer to
the $R_{01}$ field equation as the ``trivial equation.''

The structure of (\ref{eq:35})--(\ref{eq:37}) provides insight into
how the field equations propagate initial data given on a lightlike
hypersurface.  Let us (arbitrarily) single out $u^0$ as ``time,'' and
suppose that the six ``evolutionary'' vacuum equations
\begin{equation}
^{(4)}\!R_{ab}=0,\qquad R_{00}=R_{0a}=0
\label{eq:38}
\end{equation}
are satisfied everywhere in the neighbourhood of a hypersurface
 $u^0={\rm const}$. (Bondi and
Sachs refer to $^{(4)}\!R_{ab}$ as the ``main equations''
and to $R_{00}$, $R_{0a}$ as ``hypersurface equations.''
$R_{00}$ is, in fact, the Raychaudhuri focusing equation \cite{ref:11},
governing the expansion of the lightlike normal
$\ell_{(0)}=-\ell^{(1)}$ to the {\it transverse\/} hypersurface
$u^1={\rm const}.$, and $R_{0a}$ similarly governs its
shear.)

Then (\ref{eq:35}) shows that the trivial equation $R_{01}=0$ is
satisfied automatically. From (\ref{eq:36}) and (\ref{eq:37}) it can be
further inferred that if $R_{11}$ and $R_{1a}$ vanish on
one hypersurface $u^0={\rm const}.$, then they will vanish
everywhere as a consequence of the six evolutionary
equations (\ref{eq:38}). This is the content of the Bondi-Sachs
lemma \cite{ref:12, ref:5}, which identifies the three conditions
$R_{11}=R_{1a}=0$ as constraints---on the expansion and
shear of the generators $\ell_{(1)}$ of an initial
hypersurface $u^0={\rm const.}$---which are respected by
the evolution.

\section{Co-ordinate conditions and gauge-fixing}
The characteristic initial-value problem \cite{ref:5} involves
specifying initial data on a given pair of lightlike
hypersurfaces $\Sigma^0$, $\Sigma^1$ intersecting in a
2-surface $S_0$.

It is natural to choose our parameters $u^A$ so that
$u^0=0$ on $\Sigma^0$ and $u^1=0$ on $\Sigma^1$. The
requirement (\ref{eq:2}) that $u^A$ be {\it globally\/} lightlike
already imposes two co-ordinate conditions on
$(u^A,\theta^a)$, considered as co-ordinates of spacetime.
Two further global conditions may be imposed. We may, for
instance, demand that $\theta^a$ be convected (Lie-propagated)
along the lightlike curves tangent to $\ell_{(0)}^\alpha$ from
values assigned arbitrarily on $\Sigma^0$. According to (\ref{eq:9}),
 this means the corresponding shift vector is zero
everywhere:
\begin{equation}
s_{(0)}^a=-\ell_{(0)}^\alpha\,\partial_\alpha\theta^a=0.
\label{eq:39}
\end{equation}
In this case, (\ref{eq:20}) shows that
\begin{equation}
\omega^a=\partial_0 s_1^a
\label{eq:40}
\end{equation}
is just the ``time''-derivative of the single remaining
shift vector.

These global co-ordinate conditions can still be
supplemented by appropriate initial conditions. We are
still free to require that $\theta^a$ be convected along
generators of $\Sigma^0$ from assigned values on $S_0$; then
\begin{equation}
s_1^a=0\qquad (u^0=0)
\label{eq:41}
\end{equation}
in addition to (\ref{eq:39}).

In addition to (or independently of) (\ref{eq:39}) and (\ref{eq:41}),
we are free to choose $u^1$ along $\Sigma^0$ and $u^0$ along
$\Sigma^1$ to be {\it affine\/} parameters of their generators.  On
$\Sigma^0$, for instance, this means, by virtue of (\ref{eq:9}) and
(\ref{eq:4}),
\[
  \ell_{(1)}^\alpha=\left(\frac{dx^\alpha}{du^1}\right)_{{\rm gen.}}
  =-g^{\alpha\beta}\partial_\beta
  u^0=-e^{-\lambda} g^{\alpha\beta} \ell_\beta^{(0)}
\]
so that $\lambda$ vanishes over $\Sigma^0$. There is a similar
argument for $\Sigma^1$. Thus, we can arrange
\begin{equation}
\lambda=0\qquad (\Sigma^0\quad{\rm and}\quad \Sigma^1).
\label{eq:42}
\end{equation}

Alternatively, in place of (\ref{eq:42}), the co-ordinate condition
\begin{equation}
D_1\lambda=\frac12K_1\quad{\rm on}\quad \Sigma^0
\label{eq:43}
\end{equation}
could be imposed to normalize $u^1$. (A corresponding
condition on $\Sigma^1$ would normalize $u^0$.) The
Raychaudhuri equation (\ref{eq:28}) for $R_{11}$ on $\Sigma^0$ would
then become linear in the expansion rate $\overline{K}_1=\partial_1\ln
g^{\frac12}$, and that facilitates its integration (cf Hayward
\cite{ref:3}, Brady and Chambers \cite{ref:7}).

\section{Characteristic initial-value problem}
We are now ready to address the question of what initial
data are needed to prescribe a unique vacuum solution of
the Einstein equations in a neighbourhood of two lightlike
hypersurfaces $\Sigma^0$ and $\Sigma^1$ intersecting in a
2-surface $S_0$ \cite{ref:5}.

We arbitrarily designate $u^0$ as ``time,'' and shall refer
to $\Sigma^0$ ($u^0=0$) as the ``initial'' hypersurface and to
$\Sigma^1$ ($u^1=0$) as the ``boundary.''

We impose the co-ordinate conditions (\ref{eq:39}), (\ref{eq:41}) and
(\ref{eq:42}) to tie down $\theta^a$ and $u^A$.  While (\ref{eq:39})
and (\ref{eq:41}) control the way $\theta^a$ are carried off $S_0$,
onto $\Sigma^0$ and into spacetime, the choice of $\theta^a$ on $S_0$
itself is unrestricted.  Thus, our procedure retains covariance under
the group of two-dimensional transformations
$\theta^a\to\theta^{a'}=f^a(\theta^b)$.

In the 4-metric $g_{\alpha\beta}$, given by (\ref{eq:12}), the following
six functions of four variables are then left undetermined:
\begin{equation}
g_{ab},\ \ \lambda,\ \ s_1^a.
\label{eq:44}
\end{equation}
(In place of $s_1^a$, it is completely equivalent to
specify $\omega^a=\partial_0 s_1^a$, since the ``initial'' value of
$s_1^a$ is pegged by (\ref{eq:41}).)

We shall formally verify that a vacuum 4-metric is uniquely
determined by the following initial data:
\renewcommand{\theenumi}{(\alph{enumi})}
\begin{enumerate}
\item On $S_0$, seven functions of two variables
$\theta^a$:
\begin{equation}
g_{ab},\ \ \omega^a,\ \ \overline{K}_A=\partial_A\,\ln g^{\frac12}\qquad
(S_0);
\label{eq:45}
\end{equation}
\item on $\Sigma^0$ and $\Sigma^1$, two independent
functions of three variables which specify the intrinsic
conformal 2-metric:
\begin{equation}
g^{-\frac12}\,g_{ab}\qquad (\Sigma^0\ {\rm  and }\ \Sigma^1).
\label{eq:46}
\end{equation}
\end{enumerate}

Instead of (\ref{eq:46}), it is equivalent to give the shear rates
of the respective generators,
\begin{equation}
{\sigma_1}_a^b\ {\rm  on }\ \Sigma^0,\quad {\sigma_0}_a^b\ {\rm  on
}\ \Sigma^1,
\label{eq:47}
\end{equation}
defined as the trace-free extrinsic curvatures:
\begin{equation}
\sigma_{Aab}=\overline{K}_{Aab}-\frac12\,g_{ab}\overline{K}_A=\frac12
g^{\frac12}\partial_A(g^{-\frac12}\,g_{ab}).
\label{eq:48}
\end{equation}
These two functions correspond to the physical degrees of
freedom (``radiation modes'') of the gravitational field
\cite{ref:5, ref:13}.

To build a vacuum solution from the initial data (\ref{eq:45}),
(\ref{eq:47}), we begin by noting that (\ref{eq:39}) implies that
$D_0=\partial_0$, $K_{0\,ab}=\overline{K}_{0\,ab}$ everywhere. Hence
the general expression (\ref{eq:28}) for $R_{00}$ reduces here to
\begin{equation}
  -R_{00}=\left(\partial_0+\frac12K_0-\lambda_{,0}\right)
   K_0+\sigma_{0ab}\sigma_0^{ab}.
  \label{eq:49}
\end{equation}
On $\Sigma^1$, we have $\lambda=\lambda_{,0}=0$ by (\ref{eq:42}).
Thus, (\ref{eq:49}) becomes an ordinary differential equation for

\[
K_0=\overline{K}_0=\partial_0\ln g^{\frac12}
\]
as a function of $u^0$. This can be integrated along the
generators, using the given data for ${\sigma_0}_a^b$ on
$\Sigma^1$, and the initial value of $\overline{K}_0$ on $S_0$, to
obtain $g^{\frac12}$, hence the full 2-metric $g_{ab}$ (hence
also $K_{0\,ab}$) over $\Sigma^1$.

Expression (\ref{eq:29}) for $R_{0a}$ reduces similarly to
\begin{equation}
	R_{0a}=-\frac12 e^{-\lambda}\left(\partial_0+
	K_0-\lambda_{,0}\right)\omega_a-\frac12(\partial_0-K_0)
	\lambda_{,a}+K_{0a;b}^{\;\;b}-\partial_a K_0
	\label{eq:50}
\end{equation}
in a spacetime neighbourhood of $\Sigma^0$ and $\Sigma^1$. On
$\Sigma^1$, since $K_{0\,ab}$ is now known, and
$\lambda=\lambda_{,a}=\lambda_{,0}=0$, (\ref{eq:50}) is a linear ordinary
differential equation for $\omega_a$ which may be integrated
along generators, with initial condition (\ref{eq:45}), to find
$\omega_a$ (hence $s_1^a$).

Thus, our knowledge of the six metric functions (\ref{eq:44}) has
been extended to all of $\Sigma^1$ with the aid of the
evolutionary equations $R_{00}=R_{0a}=0$.

A similar procedure, applied to the constraint equations
$R_{11}=R_{1a}=0$, determines the functions (\ref{eq:44}) (hence
also $K_{1ab}$) over the initial hypersurface $\Sigma^0$.
(Here we exploit (\ref{eq:41})---implying $D_1=\partial_1$%
---which holds on $\Sigma^0$ only. This limitation
is of little practical consequence, since the Bianchi
identities (Sec.~7) relieve us of the need to recheck the
constraints off $\Sigma^0$.)

Thus, the data (\ref{eq:44}), together with their tangential
derivatives $\partial_1$, $\partial_a$---which we denote shematically
by
\begin{equation}
{\cal D}=\{g_{ab},\,\lambda,\,\omega_a,\,s_1^a,\,\partial_1,\,\partial_a\}
\label{eq:51}
\end{equation}
---are now known all over the initial hypersurface
$\Sigma^0\:u^0=0$. (Note that ${\cal D}$ includes $K_{1ab}$.)

We now proceed recursively. Suppose that ${\cal D}$ is known
over some hypersurface $\Sigma:u^0={\rm const.}$ We show
that the six evolutionary equations
$^{(4)}\!R_{ab}=0$, $R_{00}=R_{0a}=0$, together with the
known boundary values of $g_{ab}$, $K_{0ab}$, $\omega_a$ and
$s_1^a$ on $\Sigma^1$, determine all first-order
time-derivatives $\partial_0$ of ${\cal D}$, and hence the
complete evolution of ${\cal D}$.

Expression (\ref{eq:27}) for the evolutionary equations
$^{(4)}\!R_{ab}=0$ can be written more explicitly, with the
aid of the identity
\begin{equation}
D_A K_{ab}^A-2K_{A(a}^{\;\;\;d}  K_{b)d}^A=-2D_1K_{0ab}+
4K_{0(a}^{\;\;\;d} K_{1b)d}+\omega_{(a;b)},
\label{eq:52}
\end{equation}
which is rooted in the symmetry
\begin{equation}
\partial_{[B}\overline{K}_{A]ab}=0,\qquad
\overline{K}_{Aab}\equiv K_{Aab}+s_{A(a;b)}
\label{eq:53}
\end{equation}
(see (\ref{eq:23}) and Appendix~B).

The equations $^{(4)}\!R_{ab}=0$ are thus seen to reduce
to a system of three linear ordinary differential equations
for $K_{0ab}$ as functions of $u^1$ on $\Sigma$, whose
coefficients are concomitants of the known data ${\cal D}$ on
$\Sigma$. Together with the boundary conditions on $K_{0ab}$
at $u^1=0$ (i.e., the intersection of $\Sigma$ with $\Sigma^1$),
they determine a unique solution for $K_{0ab}$ on $\Sigma$.

We next turn to (\ref{eq:49}) and (\ref{eq:50}) to read off the values of
$\partial_0\lambda$ and $\partial_0\omega_a$ on $\Sigma$. Since the remaining
time-derivatives are known trivially from
\[
\partial_0 s_1^a=\omega^a,\qquad \frac12\partial_0 g_{ab}=
\overline{K}_{0ab}=K_{0ab},
\]
we are now in possession of the first time-derivatives of
all the data ${\cal D}$ on $\Sigma$.

This completes our formal demonstration that the initial conditions
(\ref{eq:45}) and (\ref{eq:46}), or (\ref{eq:45}) and (\ref{eq:47}),
determine (at least locally) a unique vacuum spacetime.

\section{Lagrangian}
According to (\ref{eq:12}) and (\ref{eq:31}), the Einstein-Hilbert Lagrangian
density ${\cal L}=(-\,^4g)^{\frac12} R_\alpha^\alpha$ decomposes as
\begin{equation}
{\cal L}=g^{\frac12}e^\lambda (e^{-\lambda}R_A^A+\,^{(4)}\!R_a^a),
\label{eq:54}
\end{equation}
in which $g^{\frac12}$ refers to the determinant of $g_{ab}$.
Substitution from (\ref{eq:27}) and (\ref{eq:28}) yields the explicit form
\begin{eqnarray}
\lefteqn{
g^{-\frac12}{\cal L} =e^\lambda\, ^{(2)}\!R-D_A(2K^A+D^A\lambda)-K_A K^A-
K_A^{ab} K_{ab}^A} \nonumber \\
& & +\frac12 e^{-\lambda} \omega^a\omega_a-
e^\lambda\left(2\lambda^{;a}\,_a+\frac32 \lambda
_{,a}\lambda^{,a}\right).
\label{eq:55}
\end{eqnarray}

Second derivatives of the metric in (\ref{eq:55}) can be isolated in the
form
of a pure divergence by calling on the identities
\begin{eqnarray}
  g^{\frac12}D_A X^A&=&\partial_\alpha
  \left[(-\,^4g)^{\frac12}e^{-\lambda} X^A \ell_{(A)}^\alpha\right]-
  g^{\frac12}X^A K_A,
  \label{eq:56}\\
A^a\,_{;a}+A^a\lambda_{,a} &=&\nabla_\alpha(A^a e_{(a)}^\alpha),
\label{eq:57}
\end{eqnarray}
which follow from (\ref{eq:102}) below, and hold for any scalars $X^A$
and 2-vector $A^a$.  We thus obtain
\begin{eqnarray}
\lefteqn{
{\cal L} = -\partial_\alpha
 \left[\left(-\,^4g\right)^{\frac12} e^\lambda
 \left(2K^A+D^A\lambda\right) \ell_{(A)}^\alpha
  +2\left(-\,^4 g\right)^{\frac12}\lambda^{,a}
  e_{(a)}^\alpha\right]} \nonumber \\
  & & + g^{\frac12}\left[e^\lambda\, ^{(2)}\!R+ K_A K^A-K_A^{ab} K_{ab}^A
  +\frac12 e^{-\lambda}\omega^a\omega_a+K^A D_A\lambda+\frac12
 e^\lambda \lambda_{,a}\lambda^{,a}
\right].
\label{eq:58}
\end{eqnarray}
The divergence term integrates as usual to a surface term in the
action $S=\int {\cal L}\,d^4x$, and has no influence on the classical
equations of motion.

Variation of $S$ with respect to
\[
-e^\lambda=g_{(0)(1)}\equiv \ell_{(0)}\cdot\ell_{(1)}
\]
reproduces the
expression obtained from (\ref{eq:27}) for $G^{01}=\frac12e^\lambda R_a^a$.
Similarly,
variation with respect to $s_A^a$ yields the expression
(\ref{eq:29}) for $G_{Aa}=R_{Aa}$, if we take account of the implicit
dependence of $K_{Aab}$, $D_A$ and $\omega^a$ on $s_A^a$ through
\begin{equation}
K_{Aab}=\frac12\partial_A g_{ab}-s_{A(a;b)},
\label{eq:59}
\end{equation}
(\ref{eq:22}) and (\ref{eq:20}).  Finally, variation with respect to
$g_{ab}$ yields $^{(4)}\!G^{ab}$, if we note the identity
\begin{equation}
g^{-\frac12}\frac{\delta}{\delta g^{ab}}\int\varphi\, ^{(2)}\!R g^{\frac12}
\,d^2\theta=
g_{ab}\,\varphi^{;c}\,_c-\varphi_{;ab}.
\label{eq:60}
\end{equation}

Thus, variation of the action (\ref{eq:58}) yields eight of the ten Einstein
equations. The remaining two equations---the Raychaudhuri equations
for $R_{00}$ and $R_{11}$---cannot be retrieved directly from (\ref{eq:58}),
because the a priori conditions $\eta^{00}=\eta^{11}=0$ (expressing
the lightlike character of $u^0$ and $u^1$) which is built into
(\ref{eq:58}),
precludes us from varying with respect to these ``variables.''
 The two Raychaudhuri equations can, however, be effectively
recovered
from the other eight equations via the Bianchi idendities.

The Hamiltonian formulation of the dynamics has been discussed in
detail by Torre \cite{ref:3}. We hope to pursue this topic elsewhere.

\section{Gauss-Wein\-gar\-ten (first order) relations}
In this second half of the paper, we return to the beginning and to
the task of laying a more complete geometrical foundation for the
Ricci and Bianchi formulas which we quoted without derivation in
(\ref{eq:27})--(\ref{eq:29}) and (\ref{eq:32}), (\ref{eq:33}).  We
begin with the first-order imbedding relations for the 2-surface $S$
as a subspace of spacetime.

(\ref{eq:15}) and (\ref{eq:17}) allow us to decompose the 4-vector
$\delta e_{(a)}^\alpha /\delta\theta^b$ in terms of the basis
$(\ell_{(A)},e_{(a)})$.  Recalling (\ref{eq:5}), we find
\begin{equation}
\frac{\delta e_{(a)}}{\delta\theta^b}=-e^{-\lambda}K_{ab}^A \ell_{(A)}
+\Gamma_{ab}^c
e_{(c)}.
\label{eq:61}
\end{equation}
Similarly, in view of (\ref{eq:16}), we may decompose
\begin{equation}
\frac{\delta \ell_{(A)}}{\delta\theta^a}=L_{ABa}\ell^{(B)}+K_{Aab} e_{(b)}
\label{eq:62}
\end{equation}
where the first coefficient is given by
\begin{equation}
L_{ABa}=e^{-\lambda}\ell_{(B)}\cdot\delta\ell_{(A)}/\delta\theta^a.
\label{eq:63}
\end{equation}

This coefficient can be reduced to a much simpler form. Its
symmetric
part is
\begin{equation}
L_{(AB)a}=\frac12\,e^{-\lambda}\partial_a(\ell_{(A)}\cdot\ell_{(B)})=\frac12
\eta_{AB}\partial_a\lambda.
\label{eq:64}
\end{equation}
To obtain the skew part, we note first that
\begin{equation}
e_{(a)}^\alpha\,\ell_{[(B)}^\beta\ell_{(A)][\alpha\mid\\beta]}=0.
\label{eq:65}
\end{equation}
since $\ell_{(A)}$ is proportional to a lightlike gradient
(see (\ref{eq:4})).
With the aid of (\ref{eq:65}) and (\ref{eq:19}) we now easily derive
\begin{eqnarray}
L_{[AB]a} &=&e^{-\lambda}\ell_{[(B)}^\beta\ell_{(A)]\beta\mid\alpha}
e_{(a)}^\alpha=\frac12
e^{-\lambda}[\ell_{(B)},\ell_{(A)}]^\alpha\, e_{(a)\alpha}\nonumber \\
&=&\frac12 \epsilon_{AB}\omega_a e^{-\lambda},
\label{eq:66}
\end{eqnarray}

Combining (\ref{eq:64}) and (\ref{eq:66}), we arrive at the simple expression
\begin{equation}
2L_{ABa}=\eta_{AB}\partial_a\lambda+e^{-\lambda}\epsilon_{AB}\omega_a
\label{eq:67}
\end{equation}
for the first coefficient in (\ref{eq:62}).

The Gauss-Weingarten equations (\ref{eq:61}) and (\ref{eq:62}) govern
the variation of
%--- included an and
the 4-vectors $\ell_{(A)}$ and $\,e_{(a)}$ along directions tangent to
$S$. We now turn to their variation along the two normals.

We have from (\ref{eq:4}),
\begin{equation}
	\nabla_\beta\ell_{(A)\alpha}=
	2\ell_{(A)[\alpha}\partial_{\beta]}\lambda+\nabla_\alpha\,\ell_{
	(A)\beta}.
	\label{eq:68}
\end{equation}
Multiplying by $\ell_{(B)}^\beta$ and symmetrizing in $A$, $B$ gives
\begin{equation}
(\ell_{(B)}\cdot\nabla)\ell_{(A)}+(\ell_{(A)}\cdot\nabla)\ell_{(B)}
=2\ell_{((A)}D_{B)}\lambda-\eta_{AB}e^\lambda\nabla\lambda
\label{eq:69}
\end{equation}
where $D_A\lambda$ is defined as in (\ref{eq:22}).
 It follows that
\begin{equation}
\nabla\lambda=e^{-\lambda}\ell^{(A)} D_A\lambda+e^{(a)}\partial_a\lambda.
\label{eq:70}
\end{equation}
On the other hand, the {\it difference\/} of the two terms on the left
of (\ref{eq:69}) is given by (\ref{eq:19}) as $\epsilon_{AB}\omega_a
e^{(a)}$.  Adding finally yields
\begin{equation}
  (\ell_{(B)}\cdot\nabla)\ell_{(A)}=N_{ABC}\ell^{(C)}-e^\lambda
  L_{BAa}e^{(a)},
\label{eq:71}
\end{equation}
where
\begin{equation}
N_{ABC}=D_{(A}\lambda\eta_{B)C}-\frac12\eta_{AB}D_C\lambda,
\label{eq:72}
\end{equation}
and $L$ was defined in (\ref{eq:67}).

Proceeding finally to the transverse variation of $e_{(a)}$, we
have
from (\ref{eq:14}) and (\ref{eq:9}),
\[
  \frac{\delta e_{(a)}}{\delta u^A}=\frac{\delta}{\delta\theta^a}
  \left(\ell_{(A)}+s_A^b e_{(b)}\right).
\]
Substituting from (\ref{eq:61}) and (\ref{eq:62}), it is
straightforward to reduce this to
\begin{equation}
\frac{\delta e_{(a)}}{\delta u^A}=(L_{ABa}-s_A^b e^{-\lambda}K_{Bab})
\ell^{(B)}+\widetilde{K}_{Aa}\,^b\,e_{(b)}
\label{eq:73}
\end{equation}
where
\begin{equation}
\widetilde{K}_{Aab}=K_{Aab}+s_{Ab;a}.
\label{eq:74}
\end{equation}

Applying Leibnitz's rule to
\[
\partial_A g_{ab}=\frac{\delta}{\delta u^A}(e_{(a)}\cdot e_{(b)}),
\]
we read off from (\ref{eq:73}) the result
\begin{equation}
\frac12\partial_A g_{ab}=\overline{K}_{Aab}\equiv\widetilde{K}_{A(ab)},
\label{eq:75}
\end{equation}
which gives direct geometrical meaning to the extrinsic curvature
in
terms of transverse variation of the 2-metric.

The normal absolute derivatives of $e_{(a)}$ are given by
(recalling
(\ref{eq:9}) and (\ref{eq:13}))
\[
(\ell_{(A)}\cdot\nabla)e_{(a)}=\frac{\delta e_{(a)}}{\delta
u^A}-s_A\,^b\,\frac{\delta e_{(a)}}{\delta\theta^b}.
\]
With the help of (\ref{eq:73}) and (\ref{eq:61}) this reduces to
\begin{equation}
(\ell_{(A)}\cdot\nabla)e_{(a)}=L_{ABa}\ell^{(B)}+(
\widetilde{K}_{Aa}\,^b-s_A\,^c\Gamma_{ac}^b)e_{(b)}
\label{eq:76}
\end{equation}
Correspondingly, the two normal derivatives of $g_{ab}$ are
\begin{equation}
(\ell_{(A)}\cdot\nabla)g_{ab}=2\overline{K}_{Aab}-s_A^c\partial_c
g_{ab}.
\label{eq:77}
\end{equation}

The two-dimensionally noncovariant terms which appear in (\ref{eq:76}) and
(\ref{eq:77}) are not a mistake. They arise because the normal gradient
$\ell_{(A)}\cdot\nabla$, applied to objects carrying lower-case Latin
indices---let us say $g_{ab}$---does not preserve manifest
two-dimensional covariance, since it contains (see (\ref{eq:22})) a piece
$-s_A^c\partial_c g_{ab}$ involving ordinary (rather than
two-dimensional
covariant) derivatives with respect to $\theta^c$. Although not
incorrect, this is a formal impediment: it threatens to
clutter
our formulae with terms in the shift vectors $s_A^a$ which are, to
boot, noncovariant. In the following section, we explain how this
can
be remedied by introducing a ``rationalized'' gradient operator
$\widetilde{\nabla}$.

\section{Rationalized operators $\widetilde{\nabla}$,
 $\widetilde{D}_A$, $\nabla_a$}
The rationalized operator $\widetilde{\nabla}_\alpha$ avoids the
two-dimensionally noncovariant terms which appear when $\nabla_\alpha$
is applied to objects bearing lower-case Latin indices, as in
(\ref{eq:61}), (\ref{eq:76}) and (\ref{eq:77}).

Applied to scalar fields or to 4-tensors not bearing lower-case
Latin
indices, $\widetilde{\nabla}$ is identical with $\nabla$. If the object does
carry
such indices, there are supplementary terms involving the
two-dimensional connection $\Gamma_{bc}^a$.

Specifically, we define
\begin{equation}
  \widetilde{\nabla}_\alpha=\nabla_\alpha+p_\alpha^{(a)}
  (\nabla_a-e_{(a)}\cdot\nabla)
  \label{eq:78}
\end{equation}
in which $e_{(a)}\cdot\nabla\equiv\delta/\delta\theta^a$ is the absolute
derivative introduced in (\ref{eq:13}), and the operator $\nabla_a$ will be
specified in a moment. We have introduced the pair of 4-vectors
$p^{(a)}=\nabla\theta^a$, i.e.,
\begin{equation}
p_\alpha^{(a)}=\partial\theta^a/\partial x^\alpha.
\label{eq:79}
\end{equation}
Their projections onto the basis vectors are, according to (\ref{eq:6}) and
(\ref{eq:9}),
\begin{equation}
p^{(a)}\cdot e_{(b)}=\delta_b^a,\qquad p^{(a)}\cdot\ell_{(A)}=-s_A^a,
\label{eq:80}
\end{equation}
from which follows the identity
\begin{equation}
  \delta_\alpha^\beta-p_\alpha^{(a)}e_{(a)}^\beta
  =e^{-\lambda}\ell_\alpha^{(A)} \partial x^\beta/\partial u^A.
  \label{eq:81}
\end{equation}

Hence (\ref{eq:78}) can be recast in terms of the absolute derivative
$\delta/\delta u^A$:
\begin{equation}
\widetilde{\nabla}=e^{-\lambda}\ell^{(A)}\partial/\partial u^A+p^{(a)}\nabla_a.
\label{eq:82}
\end{equation}

We next introduce the differential operator
\begin{equation}
\widetilde{D}_A\equiv \ell_{(A)}\cdot\widetilde{\nabla}=
\ell_{(A)}\cdot\nabla-s_A^a(\nabla_a-\delta/\delta\theta^a).
\label{eq:83}
\end{equation}
An alternative form
\begin{equation}
\widetilde{D}_A=\delta/\delta u^A-s_A^a\nabla_a
\label{eq:84}
\end{equation}
follows from (\ref{eq:82}).

Since $e_{(a)}\cdot\widetilde{\nabla}=\nabla_a$, we can reconstruct
 $\widetilde{\nabla}$
from
(\ref{eq:83}) in yet another form:
\begin{equation}
\widetilde{\nabla}=e^{-\lambda}\ell^{(A)}\widetilde{D}_A+e^{(a)}\nabla_a.
\label{eq:85}
\end{equation}

We now specify the operator $\nabla_a$. It is defined so as to act as
a
two-dimensional covariant derivative on all lower-case Latin
indices
(including parenthesized ones), and at the same time as an absolute
derivative $\delta/\delta\theta^a$ on Greek indices. Upper-case Latin
indices
are treated as inert.

As an example,
\begin{equation}
\nabla_b e_{(a)}=\delta e_{(a)}/\delta\theta^b-\Gamma_{ab}^c e_{(c)}.
\label{eq:86}
\end{equation}

It is evident that, quite generally, the ``correction''
$\nabla_a-\delta/\delta\theta^a$ in (\ref{eq:83}) and (\ref{eq:78}) is
linear and homogeneous in the two-dimensional connection
$\Gamma_{ab}^c$.

Examples of how $\nabla_a$ and $\widetilde{D}_A$ act on scalars and
2-tensors are
\begin{equation}
\begin{array}{ccc}
  \nabla_a f =\partial_a f, & & \widetilde{D}_A f = (\partial_A-s_A^a
  \partial_a)f=D_Af,\\
  \nabla_a X_{\ c\dots}^{b\dots} = X_{c\dots;a}^{b\dots},\widetilde{D}
  &\qquad&\widetilde{D}_A X_{\ c\dots}^{b\dots}
  =(\partial_A-s_A^a\nabla_a)X_{\ c\dots}^{b\dots}.
\end{array}
\label{eq:87}
\end{equation}

For the 2-metric $g_{ab}$, we have from (\ref{eq:84}),
\begin{equation}
\widetilde{D}_A g_{ab}=\partial_A g_{ab},
\label{eq:88}
\end{equation}
since $\nabla_c g_{ab}\equiv g_{ab;c}=0$. Thus, (\ref{eq:75}) can be expressed
as
\begin{equation}
\frac12 \widetilde{D}_A g_{ab}=\overline{K}_{Aab},
\label{eq:89}
\end{equation}
which should be contrasted with (\ref{eq:77}).

Similarly, with the aid of (\ref{eq:86}) and (\ref{eq:83}), the noncovariant
expressions (\ref{eq:61}) and (\ref{eq:76}) become
\begin{eqnarray}
\nabla_b e_{(a)}&=&\nabla_a e_{(b)}=-e^{-\lambda}K_{Aab}\ell^{(A)},
\label{eq:90}\\
\widetilde{D}_A e_{(a)}&=&L_{ABa}\ell^{(B)}+\widetilde{K}_{Aab}
e^{(b)}.
\label{eq:91}
\end{eqnarray}

Quite generally, $\widetilde{D}_A$, $\nabla_a$ and
$\widetilde{\nabla}$ preserve both four-dimensional covariance and
covariance under ``rigid'' two-dimensional co-ordinate transformations
$\theta^a\to\theta^{a'}=f^a(\theta^b)$, with no dependence on $u^A$.
($u$-dependence of $f^a$ would induce ``gauge'' transformations of the
shift vectors $s_A^a$, see the remarks following (\ref{eq:26}).)

With the aid of (\ref{eq:85}), the last two results can be put together to
form the rationalized covariant derivative of $e_{(a)}$:
\begin{equation}
e^\lambda\widetilde{\nabla}_\beta
e_{(a)\alpha}=(L_{BAa}\ell_\alpha^{(A)}+
\widetilde{K}_{Bab}e_\alpha^{(b)})\ell_\beta^{(B)}
-K_{Aab}\ell_\alpha^{(A)} e_\beta^{(b)}.
\label{eq:92}
\end{equation}

Equations~(\ref{eq:62}) and (\ref{eq:71}) similarly combine to produce
\begin{equation}
\nabla_\beta\ell_{(A)\alpha}=(e^{-\lambda}N_{ABC}\ell_\alpha^{(C)}
-L_{BAa}e_\alpha^{(a)})\ell_\beta^{(B)}+(L_{ABb}\ell_
\alpha^{(B)}+K_{Aab}e_\alpha^{(a)})e_\beta^{(b)}.
\label{eq:93}
\end{equation}
(No distinction here between $\widetilde{\nabla}$ and $\nabla$, since
$\ell_{(A)}$ carries no lower-case Latin indices.)

{\it To sum up\/}: equations (\ref{eq:92}) and (\ref{eq:93})
encapsulate the full set of first-order (Gauss-Wein\-gar\-ten)
equations, which control tangential and normal variations of the basis
vectors $e_{(a)}$, $\ell_{(A)}$.  The coefficients in these equations
are given by (\ref{eq:16}), (\ref{eq:67}), (\ref{eq:20}),
(\ref{eq:72}) and (\ref{eq:74}).  Their geometrical meaning emerges
from (\ref{eq:75}), (\ref{eq:19}) and
remarks following those equations.

\section{Rationalized Ricci commutation rules}
The usual commutation relations need to be modified for
$\widetilde{\nabla}_\alpha$.
To derive the modified form, consider the action of $\widetilde{\nabla}$ on
any
field object $X_a$ bearing just one lower-case Latin and an
arbitrary
set of other indices.

{}From (\ref{eq:78}) and (\ref{eq:86}),
\[
\widetilde{\nabla}_\gamma\widetilde{\nabla}_\beta
 X_a=(\delta_a^c \nabla_\gamma-p_\gamma^{(n)}\Gamma_{na}^c)
 (\delta_c^b\nabla_\beta-p_\beta^{(m)}\Gamma_{mc}^b) X_b.
\]
Skew-symmetrizing with respect to $\beta$ and $\gamma$, and noting from
(\ref{eq:79}) that $\nabla_{[\gamma}p_{\beta]}^{(m)}=0$ leads to
\begin{equation}
  \widetilde{\nabla}_{[\gamma}\widetilde{\nabla}_{\beta]}X_a
  =\nabla_{[\gamma}\nabla_{\beta]}
  X_a+p_{[\beta}^{(m)}\big\{p_{\gamma]}^{(n)}\frac12
  R^b\,_{amn}-e^{-\lambda}\ell_{\gamma]}^{(A)}\partial_A\Gamma_{ma}^b\big\}X_b,
\label{eq:94}
\end{equation}
in which $\partial_\gamma\Gamma_{ma}^b$ has been expanded using
\begin{equation}
  \partial_\gamma=e^{-\lambda}\ell_\gamma^{(A)}
  \partial_A+p_\gamma^{(n)}\partial_n,
  \label{eq:95}
\end{equation}
which is a special case of (\ref{eq:82}).

The right-hand side of (\ref{eq:94}) can be further reduced:
$\partial_A\Gamma_{ma}^b$ is a 2-tensor, given by
\begin{equation}
\partial_A\Gamma_{ma}^b=2\overline{K}_{A(a;m)}^b-
\overline{K}_{Ama}\,^{;b}
\label{eq:96}
\end{equation}
according to (\ref{eq:75}); and in two dimensions we have
\begin{equation}
R^b\,_{amn}=\,^{(2)}\!R\delta_{[m}^b g_{n]a}.
\label{eq:97}
\end{equation}

If $e_{(a)}^\alpha$ is substituted for $X_a$, (\ref{eq:94}),
(\ref{eq:92}) and (\ref{eq:93}) can be used to express the projection
onto $e_{(a)}$ of the four-dimensional Riemann tensor in terms of the
first-order Gauss-Weingarten variables $K$, $L$, $N$ and their
derivatives.  If our interest is primarily in the Ricci tensor, the
contracted form $(\gamma=\alpha)$ of (\ref{eq:94}) suffices:
\begin{equation}
(\widetilde{\nabla}_\alpha
\widetilde{\nabla}_\beta-
\widetilde{\nabla}_\beta
\widetilde{\nabla}_\alpha)e_{(a)}^\alpha=e_{(a)}^\alpha
R_{\alpha\beta}-\frac12\,^{(2)}\!R p_{(a)\beta}+e^{-\lambda}(\partial_a
\overline{K}_A)\ell_\beta^{(A)}
\label{eq:98}
\end{equation}
where
\begin{equation}
\overline{K}_A\equiv\overline{K}_{Aa}^{\,a}=\partial_A\ln g^{\frac12}
\label{eq:99}
\end{equation}
and $g\equiv\det g_{ab}$.

\section{Contracted Gauss-Codazzi (second order) relations{.} Ricci
tensor}
The Gauss-Codazzi relations are the integrability conditions of the
system of first order (Gauss-Weingarten) differential equations (\ref{eq:92}),
(\ref{eq:93}). As just noted, they express projections of the four-dimensional
Riemann tensor in terms of $K$, $L$, $N$ and their first derivatives.

Contraction of these equations gives frame components of the Ricci
tensor. The most concise way of deriving these components in practice
is through recourse to a generalized form of Raychaudhuri's equation
\cite{ref:11}.

Let $A^\alpha$ be an arbitrary 4-vector (which may bear arbitrary
label indices) and $B^\alpha$ a second vector free of lower-case Latin
indices, so that $\widetilde{\nabla}_\beta B^\alpha=\nabla_\beta
B^\alpha$ and the standard commutation rules apply.  Then it is easy
to check the identity
\begin{equation}
  R_{\alpha\beta}A^\alpha B^\beta=
  \widetilde{\nabla}_\beta(A^\alpha \nabla_\alpha B^\beta)-A^\alpha
  \nabla_\alpha(\nabla_\beta B^\beta)-(\widetilde{\nabla}_\beta A^\alpha)
  (\nabla_\alpha B^\beta).
\label{eq:100}
\end{equation}

If, on the other hand, $B$ is replaced by $e_{(b)}$, then we call upon
the commutation law (\ref{eq:98}) for $\widetilde{\nabla}$, with the
result
\begin{eqnarray}
\lefteqn{
R_{\alpha\beta}A^\alpha e_{(b)}^\beta =
\widetilde{\nabla}_\beta(A^\alpha \widetilde{\nabla}_\alpha
e_{(b)}^\beta)-A^\alpha
\widetilde{\nabla}_\alpha(\widetilde{\nabla}_\beta e_{(b)}^\beta)-
(\widetilde{\nabla}_\beta
A^\alpha)(\widetilde{\nabla}_\alpha e_{(b)}^\beta)} \nonumber \\
& & +\frac12\,^{(2)}R(A\cdot p_{(b)})-
e^{-\lambda}(\partial_b\overline{K}_B)(A\cdot\ell^{(B)}).
\label{eq:101}
\end{eqnarray}

With the choices $A=\ell_{(A)}$ and $e_{(a)}$, $B=\ell_{(B)}$ we can
recover all frame components of the Ricci tensor from these equations
in tandem with (\ref{eq:92}) and (\ref{eq:93}).

Some details of these calculations are recorded in
Appendix~A.
The final results have already been listed in (\ref{eq:27})--(\ref{eq:29}).

We next turn to the contracted Bianchi identities. The
projection of (\ref{eq:30}) onto $e_{(a)}$ yields
\[
\widetilde{\nabla}_\beta(R_\alpha^\beta e_{(a)}^\alpha)-R_\alpha^\beta
\widetilde{\nabla}_{\beta}\,e_{(a)}^\alpha=\frac12 \partial_a R.
\]
The second term is evaluated with the aid of (\ref{eq:92}). In the
first term, we expand
\[
e_{(a)}^\alpha R_\alpha^\beta=R_a^b e_{(b)}^\beta+e^{-\lambda} R_a^A
\ell_{(A)}^\beta,
\]
and note the (often used) results
\begin{equation}
\widetilde{\nabla}_\beta\, e_{(b)}^\beta=
\partial_b \lambda,\qquad \nabla_\beta\, \ell_{(A)}^\beta=K_A+D_A \lambda,
\label{eq:102}
\end{equation}
which follow from (\ref{eq:92}) and (\ref{eq:93}).  The result is
(\ref{eq:32}), and (\ref{eq:33}) is obtained similarly.

\section{Riemann tensor}
We list here the tetrad components of the Riemann tensor,
obtainable from the uncontracted Ricci commutation rules
(see, e.g., (\ref{eq:94})). The notation for the tetrad components
is as in Sec.~4.
\begin{eqnarray*}
^{(4)}\!R^{ab}\,_{cd}&=&\,^{(2)}\!R\,\delta^a_{[c}\delta_{d]}^b-2e^{-\lambda}
K_{A[c}^{\;\;a}K^{Ab}_{\;\;d]}\\
R_{ABCD}&=&\frac14\,\epsilon_{AB}\,\epsilon_{CD} (2e^\lambda D^E D_E
\lambda-3\omega^a\,\omega_a+e^{2\lambda}\lambda^{,a}\lambda_{,a})\\
R_{Aabc}&=&2K_{Aa[b;c]}-K_{Aa[b}\lambda_{,c]}-e^{-\lambda}\,\epsilon_{AB}
K_{a[b}^B\omega_{c]}\\
R_{aABC}&=&\frac12\,\epsilon_{BC}\{D_A\omega_a+K_{Aab}\,\omega^b
-e^\lambda\,\epsilon_{AE}(D^E\partial_a\lambda- K^E_{ab}\lambda^{,b})-\omega_a
D_A\lambda\}\\
R^{A\ B}_{\ \;a\ \ b}&=&-D^{(A}K^{B)}_{ab}+K_{bd}^A
K_a^{Bd}+D^{(A}\lambda\,
K_{ab}^{B)}-\frac12\eta^{AB}D_E\lambda\, K_{ab}^E\\
& &\qquad -\frac14\eta^{AB}(e^{-\lambda}\omega_a\omega_b+e^\lambda
\lambda_{,a}\lambda_{,b}+2e^\lambda
\lambda_{;ab})-\frac14\epsilon^{AB}g^{\frac12}\,\epsilon_{ab}\tau.
\end{eqnarray*}
We have here defined
\[
e^{-\lambda}\tau=g^{-\frac12}\,\epsilon^{ab}\,\partial_a(e^{-\lambda}\omega_b)
\]

\section{Concluding remarks}
The $(2+2)$ double-null imbedding formalism developed in
this paper leads to simple and geometrically transparent
expressions for the Einstein field equations (\ref{eq:27})--(\ref{eq:29}) and
the Einstein-Hilbert action~(\ref{eq:96}). It should find ready
application in a variety of areas, as indicated in the
Introduction.

$(2+2)$ formalisms are certainly not new \cite{ref:3, ref:4}, but they
have languished on the relativist's back-burner. We hope
that this exposition will play a role in promoting these
verstile methods from the realm of esoterica into an
everyday working tool.

\section*{Acknowledgement}

This work was supported by the Canadian Institute for
Advanced Research and by NSERC of Canada.
\appendix
\section{Computing Ricci components: some
intermediate details}
For the convenience of enterprising readers who wish to
derive the Ricci components (\ref{eq:64})--(\ref{eq:66}) for themselves, we
record here some intermediate steps of the computations.

Computation of $^{(4)}\!R_{ab}$ from (\ref{eq:101}) requires
evaluation of
\begin{eqnarray*}
  \widetilde{\nabla}_\beta(e_{(a)}^\alpha
  \widetilde{\nabla}_\alpha e_{(b)}^\beta)
  &=&-e^{-\lambda} (\widetilde{D}_A+ K_A) K_{ab}^A\\
  (\widetilde{\nabla}_\beta e_{(a)}^\alpha)
  (\widetilde{\nabla}_\alpha e_{(b)}^\beta)
  &=&\frac12(\lambda_{,a}\lambda_{,b}+e^{-2\lambda}
  \omega_a\omega_b)-2e^{-\lambda}K_{A(a}^{\ \ d}\widetilde{K}_{\;\;b)d}^A,
\end{eqnarray*}
which can be verified from (\ref{eq:92}) and (\ref{eq:67}).

Computation of $R_{AB}$ from (\ref{eq:100}) requires
\begin{eqnarray*}
\ell_{(A)\alpha|\beta}\ell_{(B)}^{\beta|\alpha} &=&(D_A \lambda)
(D_B\lambda)-\frac12\eta_{AB}(D_E\lambda)(D^E\lambda)\\
&  & +K_{Aab}K_B^{ab}-\frac12 e^\lambda\eta_{AB}
(\lambda_{,a}\lambda^{,a}+e^{-2\lambda}\omega^a \omega_a)
\end{eqnarray*}
which follows from (\ref{eq:93}), (\ref{eq:67}) and (\ref{eq:72}).

Finally, computation of $R_{Aa}$ requires
\begin{eqnarray*}
\lefteqn{
(\widetilde{\nabla}_\beta \,e_{(a)}^\alpha)\ell_{(A)|\alpha}^\beta
= \frac12\,\lambda_{,a} D_A\lambda+\left(K_{Aab}+\frac12 \Delta K_{Aab}\right)
\lambda^{,b}}\\
& & +\frac12\,\epsilon_{AB}e^{-\lambda}(\omega_a D^B \lambda+\Delta K_{ab}^B
\,\omega^b)
\end{eqnarray*}
in which
\[
\Delta K_{Aab}\equiv\widetilde{K}_{Aab}-K_{Aab}=s_{Ab;a}.
\]

\section{The operator $D_A$: commutation rules and other
properties}
\renewcommand{\theequation}{\thesection\arabic{equation}}
\setcounter{equation}{0}
In Sec.~3 we gave two definitions---(\ref{eq:21}) and (\ref{eq:24})---for the
operator $D_A$. It is straightforward to show their
equivalence. We have
\[
[e_{(a)},e_{(b)}]=0,\qquad [\partial x/\partial u^A, e_{(b)}]=0,
\]
since the Lie bracket of two holonomic vectors vanishes (cf
(\ref{eq:14})). In combination with (\ref{eq:9}), this yields
\begin{equation}
[\ell_{(A)},e_{(b)}]=-[s_A^a e_{(a)}, e_{(b)}]=(\partial_b
s_A^a) e_{(a)}.
\label{eq:B1}
\end{equation}
Hence, for any 2-vector $X^b$,
\begin{equation}
{\cal L}_{\ell_{(A)}}(X^b e_{(b)})=\left\{(\partial_A-
{\cal L}_{s_A^a})X^b\right\} e_{(b)}
\label{eq:B2}
\end{equation}
which proves the equivalence of (\ref{eq:24}) and (\ref{eq:21}) when applied
to $X^b$.

This argument is easily extended. For instance, for the
2-metric $g_{ab}$, the definition (\ref{eq:24}) gives
\begin{eqnarray*}
D_A g_{ab}&=&e_{(a)}^\alpha \,e_{(b)}^\beta {\cal L}_{\ell_{(A)}}
g_{\alpha\beta}=2e_{(a)}^\alpha\,e_{(b)}^\beta\ell_{(A)(\alpha|\beta)}\\
&=& 2K_{Aab}
\end{eqnarray*}
by (\ref{eq:16}), which agrees with the form (\ref{eq:23}) obtained from the
definition (\ref{eq:21}).
(Strictly speaking, (\ref{eq:24}) requires that the projector
$\Delta_{\alpha\beta}\equiv e_{(a)\alpha}\,e_\beta^{(a)}$ should replace
$g_{\alpha\beta}$ in the first equality above. But, according to
the completeness relation (\ref{eq:11}), the difference involves the
Lie derivative of $\ell_\alpha^{(A)} \ell_{(A)\beta}$, which is
linear homogeneous in $\ell_{(A)}$ and projects to zero.)

Commutation relations for $D_A$ follow most easily from the
definition (\ref{eq:24}). For a scalar field $f$,
\[
2D_{[B}D_{A]} f=[{\cal L}_{\ell_{(B)}},{\cal L}_{\ell_{(A)}}]f
={\cal L}_{[\ell_{(B)},\ell_{(A)}]}f=\epsilon_{AB} \omega^a
\partial_a f,
\]
where we have recalled the well-known result that the
commutator of two Lie derivatives is the Lie derivative of
the commutator (i.e., Lie bracket), and made use of (\ref{eq:19}).

Consider next the operation on a 2-vector $X^a$. We have
from (\ref{eq:24}),
\begin{eqnarray*}
D_B D_A X^a &=&e_\beta^{(a)}{\cal L}_{\ell_{(B)}}(e_{(b)}^\beta
D_A X^b)\\
&=&e_\beta^{(a)} {\cal L}_{\ell_{(B)}}\{\Delta_\alpha^\beta
{\cal L}_{\ell_{(A)}}(e_{(a)}^\alpha X^a)\}.
\end{eqnarray*}
The projection tensor $\Delta_\alpha^\beta$ can be replaced by
$\delta_\alpha^\beta$, because the Lie derivative, operating on the
difference, gives terms proportional to $\ell_{(E)}^\beta$ or
$\ell_{(E)\alpha}$, which project to zero, noting (\ref{eq:B2}). Thus,
\[
D_B D_A X^a=e_\beta^{(a)} {\cal L}_{\ell_{(B)}}
{\cal L}_{\ell_{(A)}}(e_{(a)}^\alpha X^a).
\]
We can now proceed exactly as for the scalar case to derive
the commutator. The result (generalized to an arbitrary
2-tensor) is
\[
[D_B,D_A]X_{\ b\dots}^{a\dots}=\epsilon_{AB} {\cal L}_{\omega^d}
X_{\ b\dots}^{a\dots}.
\]

In particular,
\[
[D_B, D_A]g_{ab}=2\epsilon_{AB} \omega_{(a;b)}.
\]
Recalling (B3), this may be written
\[
D_{[B}K_{A]ab}=\frac12\,\epsilon_{AB}\,\omega_{(a;b)},
\]
which was used in (\ref{eq:52}). It contracts to
\[
D_{[B}K_{A]}=\frac12\,\epsilon_{AB}\,\omega^a_{\ ;a}.
\]

These last two identities also play a role in
symmetrizing---or, more properly, recognizing the implicit
symmetry of---the raw expressions for $R_{AB}$ and
$R_{AaBb}$ that emerge from the Ricci commutation
relations. The manifestly symmetric expressions listed in
(\ref{eq:28}) and Sec.~13 have been symmetrized with the aid of
these identities.

To conclude, we note the rule for commuting $D_A$ and the
two-dimensional covariant derivative $\nabla_a$. The
commutator $[D_A,\nabla_a]$, applied to any 2-tensor, is
formed by a pattern similar to its two-dimensional covariant
derivative, but with $\Gamma_{bc}^a$ replaced by
\[
D_A\Gamma_{bc}^a=2K_{A(b;c)}^{\ \ a}-K_{Abc}\,^{;a}.
\]
As examples:
\begin{eqnarray*}
 [ D_A,\nabla_a ] X^b & = & X^d D_A \Gamma_{da}^b, \\
 {}[ D_A,\nabla_a  g_{bc} ] & = & -2(D_A\Gamma_{a(b}^d)g_{c)d}=-2K_{Abc;a}
\end{eqnarray*}

The justification for the rule is that the partial derivative
$\partial_a$(applied to any two-dimensional geometrical object)
commutes with both $\partial_A$ and the two-dimensional Lie
derivative ${\cal L}_{s_A^d}$, sothat, by (\ref{eq:21}),
\[
 \left[ D_A, \partial_a\right] X^{b\dots}_{\ c\dots} = 0.
\]


\begin{thebibliography}{99}

\bibitem{ref:1} Arnowitt R, Deser S and Misner C W 1962 {\it
Gravitation: an Introduction to Current Research}, ed Witten
(New~York: Wiley) Chap.~7

\bibitem{ref:2} Misner C W, Thorne K S and Wheeler J A 1973 {\it
Gravitation} (San Francisco: Freeman) Chap.~21

\bibitem{ref:3} d'Inverno R A and Smallwood J 1980 Phys. Rev. D {\bf 22} 1233
\newline
Torre C G 1986 Class. Quantum Grav. {\bf 3} 773
\newline
McManus D 1992 J. Gen. Rel. Grav. {\bf 24} 65
\newline
Hayward S A 1993 Class. Quantum Grav. {\bf 10} 779
\newline
d'Inverno R A and Vickers J A G 1995 Class. Quantum Gravity {\bf 12}
753

\bibitem{ref:4} Geroch R, Held A and Penrose R 1973 J. Math. Phys. {\bf 14} 874

\bibitem{ref:5} Sachs R K 1962 J. Math. Phys. {\bf 3} 908\newline
Dautcourt G 1963 Ann. Physik {\bf 12} 202\newline
Penrose R 1980 J. Gen. Rel. Grav. {\bf 12} 225\newline
Friedrich H 1981 Proc. Roy. Soc. A {\bf 375} 169\newline
Stewart J M and Friedrich H 1982 Proc. Roy. Soc. A {\bf 384}\newline
Rendall A D 1990 Proc. Roy. Soc. A {\bf 427} 221\newline
Stewart J M 1990 {\it Advanced General Relativity} (Cambridge:
Cambridge Univ. Press) Chap.~4

\bibitem{ref:6} Unruh W G, Hayward G, Israel W and McManus D 1989 Phys. Rev.
Letters {\bf 62} 2897

\bibitem{ref:7} Israel W 1986 Phys. Rev. Letters {\bf 56} 789\newline
\hphantom{Israel W} 1986 Phys. Rev. Letters {\bf 57} 397\newline
\hphantom{Israel W} 1986 Can. J. Phys. {\bf 64} 120\newline
Hayward S A 1994 Phys. Rev. D {\bf 49} 6467\newline
\hphantom{Hayward S A} 1994 Class. Quantum Grav. {\bf 11} 3025
\newline
Brady P R and Chambers C M 1995 Phys. Rev. D {\bf 51} 4177

\bibitem{ref:8} Dirac P A M 1949 Rev. Mod. Phys. {\bf 21} 392
\newline
Rohrlich F 1971 Acta Phys. Austriaca Suppl. {
\bf 8} 277\newline
Frolov V P 1978 Fortschr. Physik {\bf 26} 455\newline
Goldberg J N 1985 Found. Physics {\bf 15} 439
\newline
Convery M E, Taylor C C and Jun J W 1995 Phys. Rev. D {\bf 51} 4445

\bibitem{ref:new9} Verlinde E and Verlinde H 1993 {\it String Quantum
  Gravity and Physics at the Planck Energy Scale} ed Sanchez (Singapore:
  World Scientific) p.  262\\
  Kallosh R 1992 Phys.  Let.  {\bf B275} 284

\bibitem{ref:9} Brady P R, Droz S, Israel W and Morsink S M  1995 Paper
submitted for publication

\bibitem{ref:10} Synge J L 1960 {\it Relativity: The General Theory}
(Amsterdam:
North-Holland) p.~1

\bibitem{ref:11} Wald R M 1984 {\it General Relativity} (Chicago: Univ. of
Chicago Press) p.~218

\bibitem{ref:12} Bondi H, van der Burg MGJ and Metzner AWK 1962 Proc.
Roy. Soc. A {\bf 269} 21

\bibitem{ref:13} d'Inverno R A and Stachel  J 1978 J. Math. Phys. {\bf 19} 2447



\end{thebibliography}
\end{document}